\documentclass[twocolumn,showpacs,amssymb]{revtex4}
\usepackage{graphicx}
\usepackage{dcolumn}
\usepackage{bm}
\usepackage{epsfig}

\begin{document}

\title{SGR-like behaviour of the repeating FRB 121102}
\author{F. Y. Wang$^{1,2\ast}$ and H. Yu$^{1}$}
\affiliation{$^{1}$School of Astronomy and Space Science, Nanjing
University, Nanjing 210093, China \\
$^{2}$Key Laboratory of Modern Astronomy and Astrophysics (Nanjing
University), Ministry of Education, Nanjing 210093, China \\
$^\ast$Electronic address: fayinwang@nju.edu.cn}

\date{\today}

\begin{abstract}
{Fast radio bursts (FRBs) are millisecond-duration radio signals
occurring at cosmological distances. However the physical model of
FRBs is mystery, many models have been proposed. Here we study the
frequency distributions of peak flux, fluence, duration and waiting
time for the repeating FRB 121102. The cumulative distributions of
peak flux, fluence and duration show power-law forms. The waiting
time distribution also shows power-law distribution, and is
consistent with a non-stationary Poisson process. These
distributions are similar as those of soft gamma repeaters (SGRs).
We also use the statistical results to test the proposed models for
FRBs. These distributions are consistent with the predictions from
avalanche models of slowly driven nonlinear dissipative systems.}
\end{abstract}


\maketitle

\section{Introduction}

Fast radio bursts (FRBs) are intense radio flashes occurring at high
Galactic latitudes with anomalously high dispersion measure (DM)
\cite{Lorimer07,Thornton13,Masui15,Katz16b,Ravi16}. Duo to the lack
of distance information, their physical origin is unknown. Some
people suggested that the high DM is dominated by the ionized
intergalactic medium, which implies that FRBs may occur at
cosmological distances.

Recently, Keane et al. (2016) claimed to discover the first FRB host
galaxy, which is an elliptical galaxy at $z = 0.492\pm0.008$
\cite{Keane16}. However, this conclusion was questioned by some
subsequent papers \cite{Williams16,Vedantham16}. More recently,
using fast-dump interferometry with the Karl G. Jansky Very Large
Array (VLA), the host galaxy of repeating FRB 121102 was discovered
\cite{Chatterjee17,Tendulkar17}. Optical imaging and spectroscopy
identify FRB 121102 a redshift of $z = 0.19273$ \cite{Tendulkar17}.
The cosmological origin of FRB 121102 is confirmed. Therefore FRBs
are promising cosmological probes. However, the physical origin of
FRBs is mysterious until now. Many theoretical models for FRBs are
proposed, including collapses of supra-massive neutron star into
black hole \cite{Falcke14,Zhang14,Ravi14}, magnetar pulse-wind
interactions \cite{Lyubarsky14}, charged black hole binary mergers
\cite{Zhang16}, giant pulse emissions from pulsars \cite{Cordes16},
giant flares from magnetars \cite{Popov13,Kulkarni14,Katz14,Pen15,
Kulkarni16}, unipolar inductor model \cite{WangJ16}, and double
neutron stars mergers \cite{Totani13}. The FRB 121102 is repeating,
which disfavors models involving cataclysmic events
\cite{Spitler16}. Additional six bursts \cite{Scholz16} and nine
bursts \cite{Chatterjee17} for FRB 121102 are detected. So there may
be two populations of FRBs \cite{Champion16,Spitler16,Li16}. Dai et
al. (2016) proposed that the repeating bursts are produced by lots
of asteroids encountering with highly magnetized pulsar
\cite{Dai16}. A neutron star-white dwarf binary model also has been
proposed for the repeating FRB 121102 \cite{Gu16}.

Until now, twenty six bursts of FRB 121102 have been observed.
However, the nine bursts discovered by VLA are not observed by
Arecibo observatory. In this paper, we investigate the frequency
distributions of peak flux, fluence, duration and waiting time for
FRB 121102. We also test the proposed models for FRBs using the
derived distributions. This paper is organized as follows. The
frequency distributions are shown in section 2. In section 3, we
test theoretical models using the statistical results. Finally, the
conclusion and discussions are given in section 4.

\section{Frequency distributions of burst parameters}

For FRB 121102, we use the parameters of eleven bursts from
\cite{Spitler16} and six bursts from \cite{Scholz16}, which are
listed in Table 1. Because the nine bursts observed by VLA in the
2.5-3.5 GHz \cite{Chatterjee17}, and these bursts are not detected
by Arecibo, only the upper limit is given. These nine bursts are not
considered in our analysis. The eleven bursts in \cite{Spitler16}
are discovered by William E. Gordon Telescope at the Arecibo
Observatory and the 7-beam Arecibo L-band Feed Array (ALFA). The
ALFA is a seven-beam receiver operating at 1.4 GHz with 0.3 GHz
bandwidth \cite{Cordes06}. The antenna gains for these beams are
different, i.e., 10.4 K Jy$^{-1}$ for the central beam at low zenith
angles and 8.2 K Jy$^{-1}$ for the other six beams \cite{Cordes06}.
Because the bursts could be detected by different beams, the
observed flux or fluence must be corrected. Only the last six bursts
are pointing to the central beam \cite{Spitler16}, so the fluxes and
fluences of other five bursts are normalized to the central beam by
multiplying a factor of $1.268$. The additional six bursts are
observed by Green Bank Telescope and the single-pixel L-Wide
receiver at Arecibo observatory \cite{Scholz16}. Therefore, the
fluxes of these bursts are intrinsic. For each bursts, Column 2
gives the peak time of each burst listed in Column 1. The peak flux
is presented in Column 3 in unit of Jy. Column 4 gives the fluence
$F$ of each burst in unit of Jy ms. The observed duration time of
burst is given in Column 5. The waiting time is given in Column 6.
The waiting time $\Delta t$ is defined as the difference of
occurring times for two adjacent bursts, and can be calculated from
the time difference of Column 2. Only the continues observation is
considered. When calculating the waiting time, the peak flux limit
0.02 Jy is considered. Because the detection threshold of ALFA is
about 0.02 Jy \cite{Spitler16,Scholz16}. The definition of waiting
time is widely used in solar physics and astrophysics.

The number of bursts $N(F)dF$ with fluence between $F$ and $F + dF$
can be expressed by
\begin{equation}\label{fluencedis}
N(F)dF\propto F^{-\alpha_F}\,dF,
\end{equation}
where $\alpha_F$ is the power-law index. The number of bursts for
FRB 121102 is small. Rather than examining the differential
distribution directly, it is preferable to plot the cumulative
distribution, which can avoid binning of the data. Because the width
of binning can affect the fitting result. Integrating equation
(\ref{fluencedis}), we obtain the cumulative distribution of fluence
\begin{equation}
N(>F)\propto\int_F^\infty F^{-\alpha_F}\,dF \propto F^{-\alpha_F+1}.
\end{equation}
For the peak flux $S$, the differential frequency distribution is
\begin{equation}\label{fluxdis}
N(S)dS\propto S^{-\alpha_S}\,dS.
\end{equation}
So the number of FRBs with peak flux larger than $S$ is
\begin{equation}
N(>S)\propto\int_S^\infty S^{-\alpha_S}\,dS \propto S^{-\alpha_S+1}.
\end{equation}

We apply the Markov Chain Monte Carlo (MCMC) method to derive the
best-fitting parameters. In astrophysical observations, count
statistics is often limited. The bursts of FRB 121102 is 17. Such
low count number does not fulfill the condition required for the
Gaussian approximation, a well approximation is the Poisson
distribution. Consider the number of observed events $N_{obs}$
following Poisson distribution, the likelihood function for MCMC
method can be expressed as
\begin{eqnarray}
\mathcal{L}(\theta) &=& \sum_i\ln(P_i(N_{obs,i})) \\ \nonumber
&=&\sum_i(N_{obs,i}\ln(N_{th}(\theta))-\ln(N_{obs,i}!)-N_{th}(\theta)),
\end{eqnarray}
where $\theta$ is the parameter in the model to be constrained by
the observed data, $N_{obs,i}$ is the $i$th observed data, and
$N_{th}$ is the theoretical number predicted by model. For the
cumulative distribution, it has $N_{obs,i}=i$. Therefore, the
likelihood can be re-expressed as
$\mathcal{L}(\theta)=\sum_i^{N_{obs,tot}}(i\ln(N_{th}(\theta))-\ln(i!)-N_{th}(\theta))$,
where $N_{obs,tot}$ is the total number of observed events. We use a
python package pymc \cite{github} to apply the MCMC method to
optimize the parameters of theoretical distributions. In the
fitting, we consider the priors of all the parameters $\theta$ as
uniform distributions in a relatively large range, because the
priors are not important when sampling enough samples with MCMC
method. We must note that the events in each bin of the differential
distribution are independent, but the number of events $N(>x)$ in
the cumulative distribution are statistically dependent.
Fortunately, we use a logarithmic binning, the fluctuations of
events for cumulative distribution in each bin, may follow
approximately the same random statistics $\sigma_{cum,i} =
\sqrt{N_{cum,i}}$ in each bin as for the differential distribution.
So the likelihood function of equation (5) may be a well
approximation. This problem has been extensively discussed in
\cite{Aschwanden15}. Figure 1 shows the cumulative distributions of
fluence (left panel) and peak flux (right panel) for seventeen
bursts of FRB 121102, respectively. The power-law index for fluence
is $\alpha_F=1.80\pm0.15$ with $1\sigma$ confidence level. The value
of $\alpha_F$ is from 1.5 to 2.2 \cite{Lu16}. While, for peak flux,
the power-law index is $\alpha_S=1.07\pm0.05$ with $1\sigma$
confidence level.

The differential distribution of duration time $W$ can be expressed
as
\begin{equation}
N(W)dW\propto W^{-\alpha_W}\,dW.
\end{equation}
So the cumulative distribution of duration time $W$ is
\begin{equation}
N(>W)\propto\int_W^{W_{max}} W^{-\alpha_W}\,dW \propto
W^{-\alpha_W+1}-W_{max}^{-\alpha_W+1},
\end{equation}
where $W_{max}$ is the maximal duration time. The Markov Chain Monte
Carlo (MCMC) method is also used to derive the best-fitting
parameters simultaneously. Left panel of Figure 2 presents the
cumulative distribution of duration for FRB 121102. From this panel,
a maximal duration time is obviously shown. The best-fitting
power-law index and maximal duration time are
$\alpha_W=1.95\pm0.32$, and $W_{max}=9.80\pm0.35$ with $1\sigma$
confidence level, respectively. It should be noted that the observed
duration time will be broadened when radio waves propagate through a
plasma. The scatter-broadening time of a pulsed signal depends on
the DM and the observing frequency, and an empirical function is
given \cite{Bhat04}. However, there is no clear evidence for scatter
broadening of FRB 121102 \cite{Spitler16}.

If the burst rate is constant, the waiting-time distribution is the
Poisson interval distribution \cite{Wheatland98}
\begin{equation}
P(\Delta t) = \lambda e^{-\lambda \Delta t}
\end{equation}
where $\Delta t$ is the interval between events, and $\lambda$ is
the burst rate. If the burst rate is time varying, the waiting time
distribution can be treated as a combination of piecewise constant
Poisson processes. Generally, for most forms of $\lambda(t)$, the
waiting time distribution can be shown as power-law form
\cite{Aschwanden10}
\begin{equation}
P(\Delta t) \propto \Delta t^{-\alpha_{WT}}.
\end{equation}
In order to avoid binning of the data, the cumulative waiting time
distribution is given by
\begin{equation}
N(>\Delta t)\propto \int_{\Delta t}^{\Delta t_{max}} \Delta
t^{-\alpha_{WT}}\,dW \propto \Delta t^{-\alpha_{WT}+1}-\Delta
t_{max}^{-\alpha_{WT}+1}.
\end{equation}
For FRB 121102, the observation is not continues
\cite{Spitler16,Scholz16}. The detailed observations by different
telescopes are shown in figure 1 of \cite{Scholz16}. Therefore, in
order to obtain reliable waiting times, we select the waiting times
during periods of continuous observation. We use the waiting times
presented in table 1 of \cite{Gu16}. There are ten waiting times
from tens to hundreds of seconds. Right panel of Figure 2 shows the
cumulative waiting time distribution of FRB 121102. The best-fitting
power-law index and maximal waiting time are
$\alpha_{WT}=1.09\pm0.05$, and $\Delta t_{max}=1020.18\pm 10.25$ s
with $1\sigma$ confidence level.

\section{Comparing with predictions of theoretical models}
The power-law distributions indicates the stochastic engine for FRB
121102. There are many models proposed to explain the properties of
FRBs. In this section, we will test theoretical models predictions
with statistical results.

Dai et al. (2016) proposed that the repeating bursts can be produced
from lots of asteroids encountering with highly magnetized pulsar
\cite{Dai16}. In order to explain observation, the diameters of
asteroids are small, i.e., $L<5$km \cite{Dai16}. From their equation
(2), the differential frequency distribution of diameter $L$ of
asteroids is predicted to $dN/dL=dN/dW\times dW/dL$, with duration
time $W$. So if the index for differential frequency distribution of
duration is $-2.0$, the value is $dN/dL\propto L^{-7/3}$. From the
observation of Sloan Digital Sky Survey, the a broken power law was
found with $dN/dL\propto L^{-4}$ for large asteroids (5-50 km) and
$dN/dL\propto L^{-2.3}$ for smaller asteroids (0.5-5 km)
\cite{Ivezic01}. The differential size distribution of small
asteroids is $dN/dL\propto L^{-2.29}$ \cite{Yoshida07}. These value
are well consistent with the model prediction.

Cordes and Wasserman (2016) suggested that FRBs originate from
Crab-like giant pulses of extragalactic neutron stars
\cite{Cordes16}. The index of peak flux cumulative distribution
$\alpha_S$ is from $1.3$ to $2.5$. The low limit is a little larger
than the best-fitting value. Lyutikov et al. (2016) argued that
FRBs, including repeating and non-repeating FRBs, are from giant
pulses of young rapidly rotating pulsars \cite{Lyutikov16}. In their
model, the intrinsic luminosity of an FRB is proportional to the
spin-down power of neutron star. So the predicted distribution of
FRB flux is $N(>S)\propto S^{-3/2}$ \cite{Lyutikov16}. From our
statistical study of 17 bursts of FRB 121102, the cumulative
distribution $N(>S)\propto S^{-1.06}$ is found, which is different
from their model prediction. So the repeating FRB 121102 may
disfavor the rotationally powered model. The distance measurement of
the repeating FRB 121102 also ruled out rotationally-powered radio
emission \cite{Lyutikov17}.

Katz (2016) proposed that FRBs are generated by magnetic energy
released in magnetar magnetospheres \cite{Katz16}. Metzger et al.
(2017) also argued that the repeating FRB 121102 is powered by
millisecond magnetar, through its rotational or magnetic energy
\cite{Metzger17}. From observational constraint, the magnetic energy
is favored \cite{Metzger17}. Generally, the soft gamma repeater
(SGR) outbursts result from the dissipation of magnetostatic energy
in the magnetosphere of magnetars. So we compare the statistical
properties of FRB 121102 and SGR 1806-20. Figure 3 shows the
differential distributions of duration (left panel), and waiting
time (right panel) for SGR 1806-20. We use the waiting time data
from \cite{Gogus00}, and duration time data from \cite{Gogus01}. The
best-fitting indices are $\alpha_W=2.14\pm0.22$ and
$\alpha_{WT}=0.95\pm0.05$ for duration time and waiting time,
respectively. The distribution of SGR 1806-20 burst energies follows
a power-law $dN\propto E^{-\gamma} dE$ with $\gamma \sim 1.6$
\cite{Cheng96,Prieskorn12}. These indices are well consistent with
those of FRB 121102. This may indicate that repeating FRBs may be
related to extremely magnetized neutron stars. Besides these
statistic distributions, there are some phenomenological similarity
between FRBs and SGRs. First, they are both repeating. At least, FRB
121102 show repeating bursts \cite{Spitler16,Scholz16,Chatterjee17}.
Second, the duty factor $D=\langle f(t) \rangle^2/\langle
f(t)^2\rangle$ with flux $f(t)$ is similar, $D\sim 10^{-10}$ for
SGRs and $D<10^{-8}$ for FRBs \cite{Law15,Katz16}. This value
denotes the the fraction of the time in which a source emits at
close to its peak flux.

\section{Discussions and Conclusion}

In this paper, we study the statistical properties of repeating FRB
121102, including peak flux, fluence, duration time and waiting
time. The cumulative distributions of peak flux, fluence and
duration show power-law forms. The waiting time distribution also
shows power-law distribution, and is consistent with a
non-stationary Poisson process. Power-law size distributions have
been discovered in many astrophysical phenomena, which may indicate
a stochastic central engine. We also compare the statistical results
with theoretical models predictions. The duration distribution from
theoretical model relating asteroids encountering with highly
magnetized pulsar is consistent with observations. Similar
distributions between FRB 121102 and SGR 1806-20, such as fluence,
duration and waiting time, also support the models proposed by
\cite{Katz16}, in which the magnetic energy releases in magnetar
magnetospheres. So more observation is needed to distinguish these
two models.

Power-law distributions of events have been discovered in a large
number of astrophysical phenomena in many wavelengths \cite[for a
recent review, see][]{Aschwanden11}. The power-law frequency
distributions, including peak flux, fluence, duration and waiting
time, are predicted by self-organized criticality (SOC) systems
\cite{Bak87,Katz86,Aschwanden11}. These distributions also satisfy
the criteria that define a SOC system \cite{Aschwanden11}, which
occurs in many natural systems that exhibit nonlinear energy
dissipation \cite{Aschwanden11,Wang13, Wang15}. Therefore, FRBs may
also be avalanche events.

In future, some facilities, such as Chinese Five-hundred-meter
Aperture Spherical radio Telescope (FAST) \cite{Nan11}, Canadian
Hydrogen Intensity Mapping Experiment (CHIME) \cite{Kaspi16}, the
Square Kilometer Array, or other upcoming wide-field telescopes,
will collect a large number of FRBs. The statistics of FRBs will
give constraints on the nature of central engine.

\section*{Acknowledgements}
We thank the anonymous referee for useful comments and suggestions.
We also thank Z. G. Dai for helpful discussion. This work is
supported by the National Basic Research Program of China (973
Program, grant No. 2014CB845800) and the National Natural Science
Foundation of China (grants 11422325 and 11373022), the Excellent
Youth Foundation of Jiangsu Province (BK20140016).

\clearpage

\begin{table}
\centering \caption{The parameters of bursts for FRB 121102.
\label{tab1}} {
\begin{tabular}{lllllllll}
    \hline
    \hline
    No. & peak time         & Peak Flux        & Fluence      & Width  & Waiting time \\
        & MJD & Jy &             Jy ms         & ms    &  s    \\
    \hline
    1 & 56233.282837008 & 0.05 & 0.13  & 3.3$\pm$0.3  &    \\
    2 & 57159.737600835 & 0.038 & 0.13  & 3.8$\pm$0.4  &     \\
    3 & 57159.744223619 & 0.038 & 0.13  & 3.3$\pm$0.4  &  572.2  \\
    4 & 57175.693143232 & 0.05 & 0.25  & 4.6$\pm$0.3  &    \\
    5 & 57175.699727826 & 0.025 & 0.11 & 8.7$\pm$1.5  &  568.9 \\
    6 & 57175.742576706 & 0.02 & 0.06 & 2.8$\pm$0.4  &  \\
    7 & 57175.742839344 & 0.02 & 0.06 & 6.1$\pm$1.4  & 22.7 \\
    8 & 57175.743510388 & 0.14 & 0.9  & 6.6$\pm$0.1  & 58.0 \\
    9 & 57175.745665832 & 0.05 & 0.3  & 6.0$\pm$0.3  & 186.2  \\
    10& 57175.747624851 & 0.05 & 0.2  & 8.0$\pm$0.5  &  169.3   \\
    11& 57175.748287265 & 0.31 & 1.0  & 3.06$\pm$0.04 & 57.2   \\
    12& 57339.356046005 & 0.04 & 0.2  & 6.73$\pm$1.12&  \\
    13& 57345.447691250 & 0.06 & 0.4  & 6.10$\pm$0.57&  \\
    14& 57345.452487925 & 0.04 & 0.2  & 6.14$\pm$1.00&  414.4  \\
    15& 57345.457595303 & 0.02 & 0.08 & 4.30$\pm$1.40&  441.3 \\
    16& 57345.462413106 & 0.09 & 0.6  & 5.97$\pm$0.35&  416.3 \\
    17& 57364.204632665 & 0.03 & 0.09 & 2.50$\pm$0.23&   \\
    \hline
    \hline
    \newline
    \end{tabular}
}

\end{table}

\clearpage

\begin{figure*}
\includegraphics[width=0.4\textwidth]{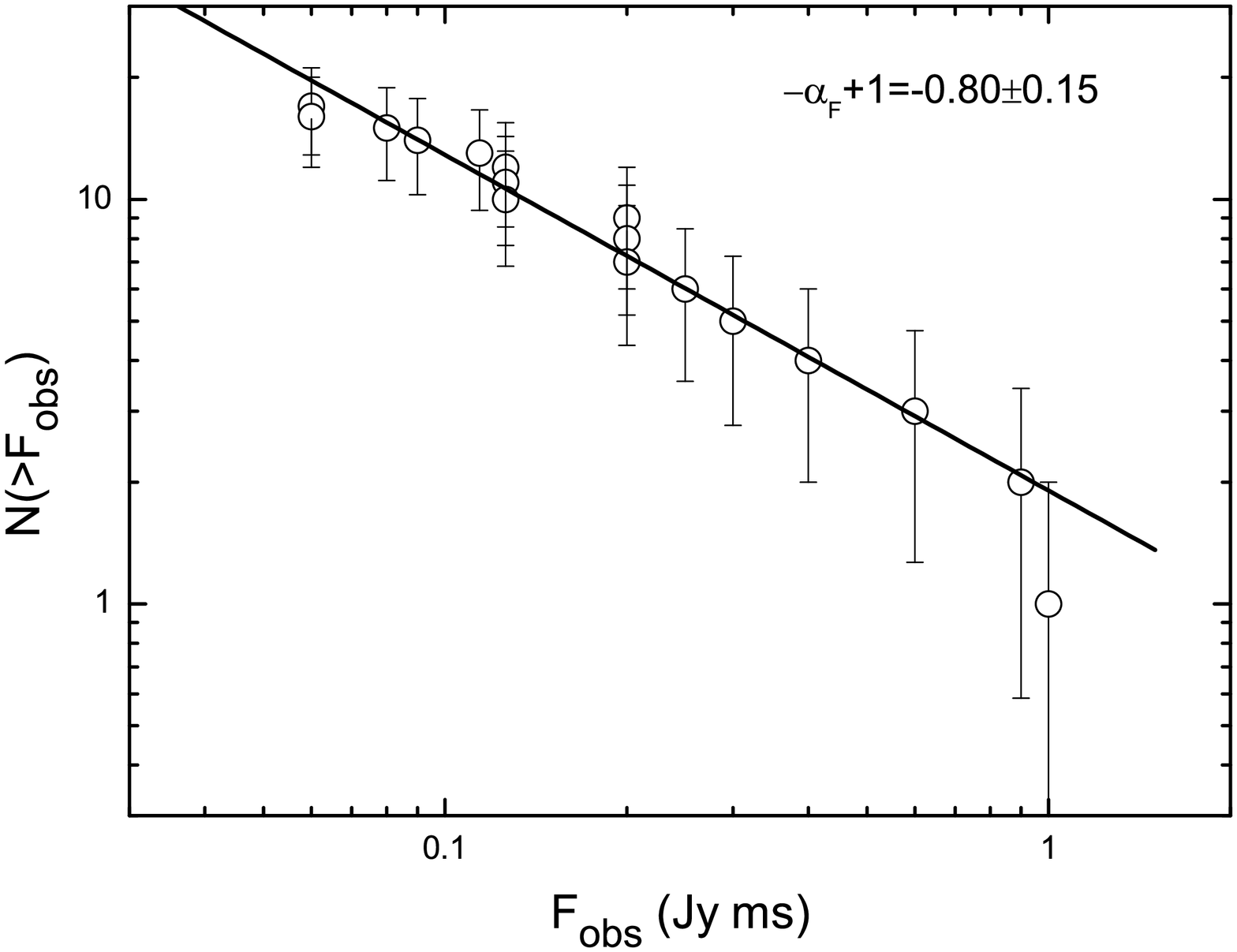}
\includegraphics[width=0.45\textwidth]{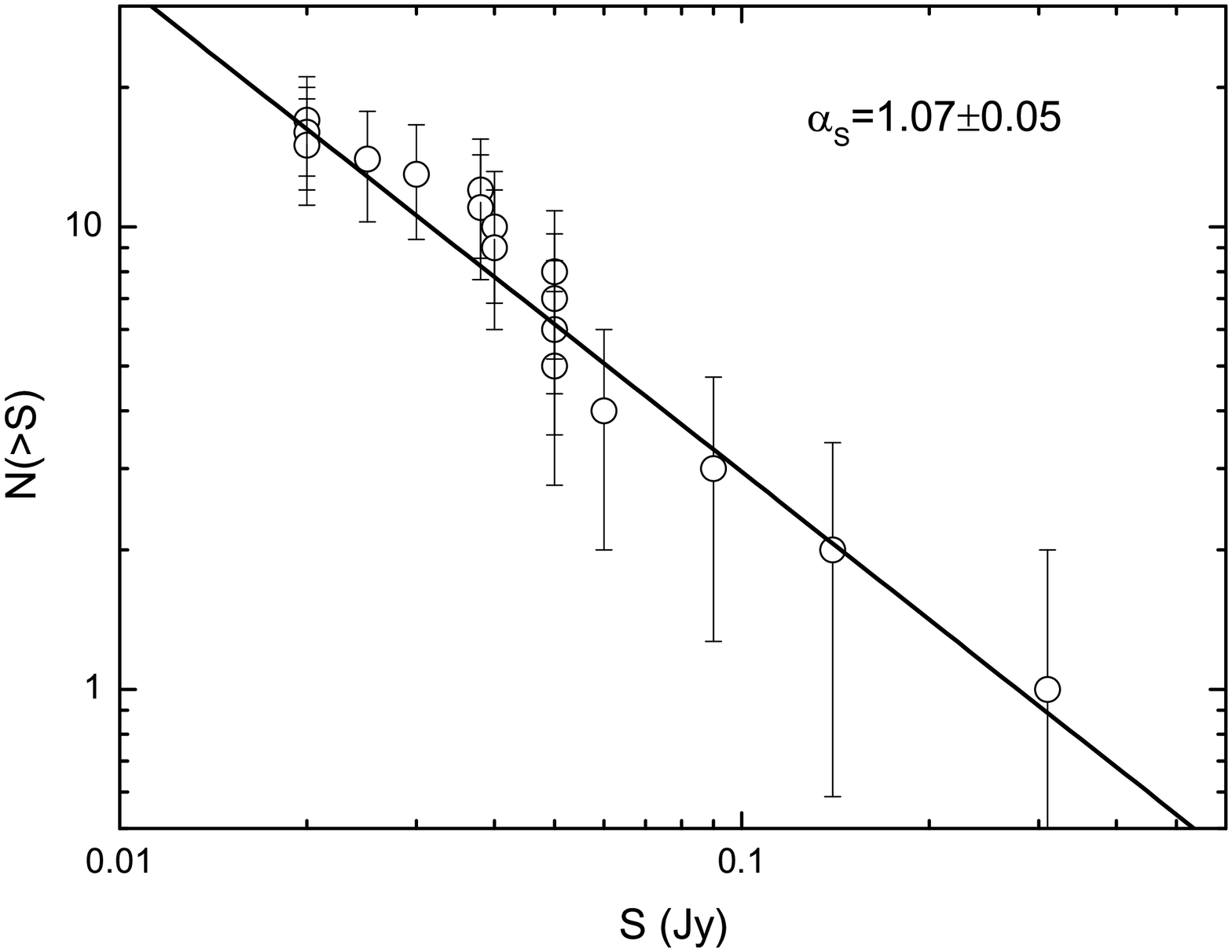}
\caption{The cumulative distributions of fluence (left panel) and
peak flux (right panel) for FRB 121102, respectively. The
best-fitting power-law indices are $\alpha_F=1.80\pm0.15$ and
$\alpha_S=1.07\pm 0.05$ for fluence and peak flux, respectively.}
\end{figure*}

\begin{figure*}
\includegraphics[width=0.4\textwidth]{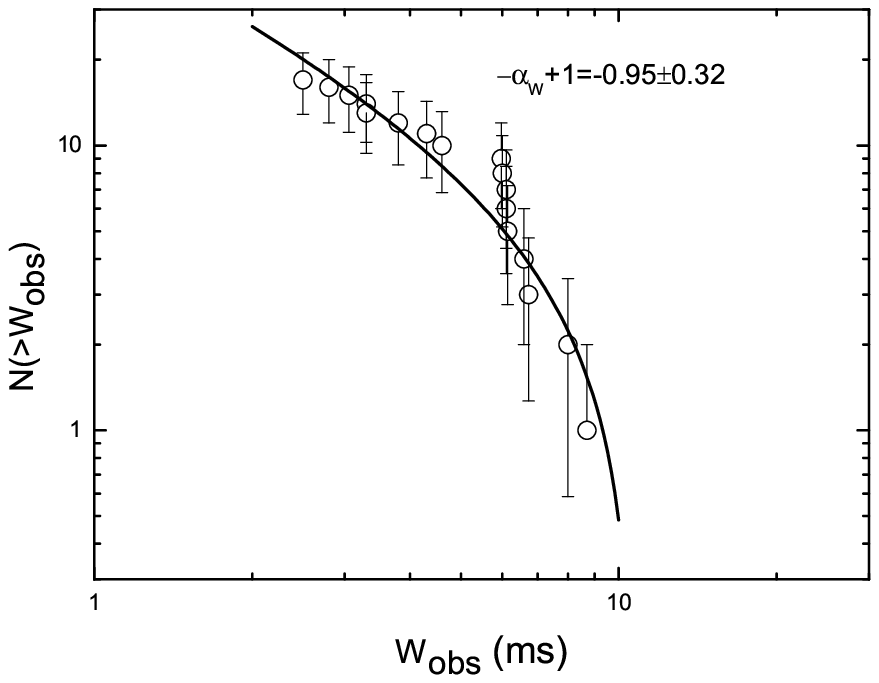}
\includegraphics[width=0.45\textwidth]{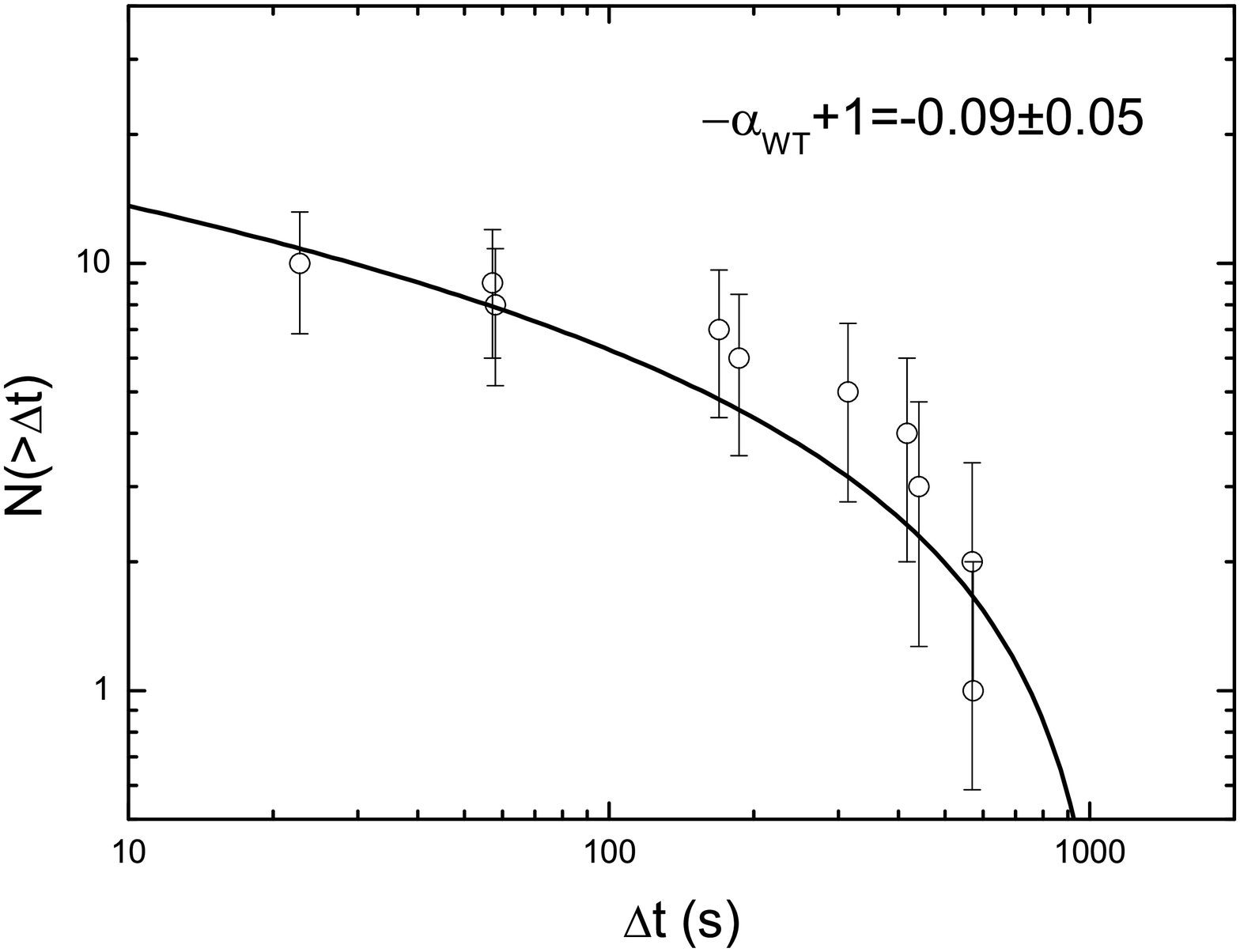}
\caption{The cumulative distributions of duration (left panel) and
waiting time (right panel) for FRB 121102, respectively. The
best-fitting power-law indices are $\alpha_W=1.95\pm0.32$ and
$\alpha_{WT}=1.09\pm0.05$, respectively.}
\end{figure*}

\begin{figure*}
\includegraphics[width=0.4\textwidth]{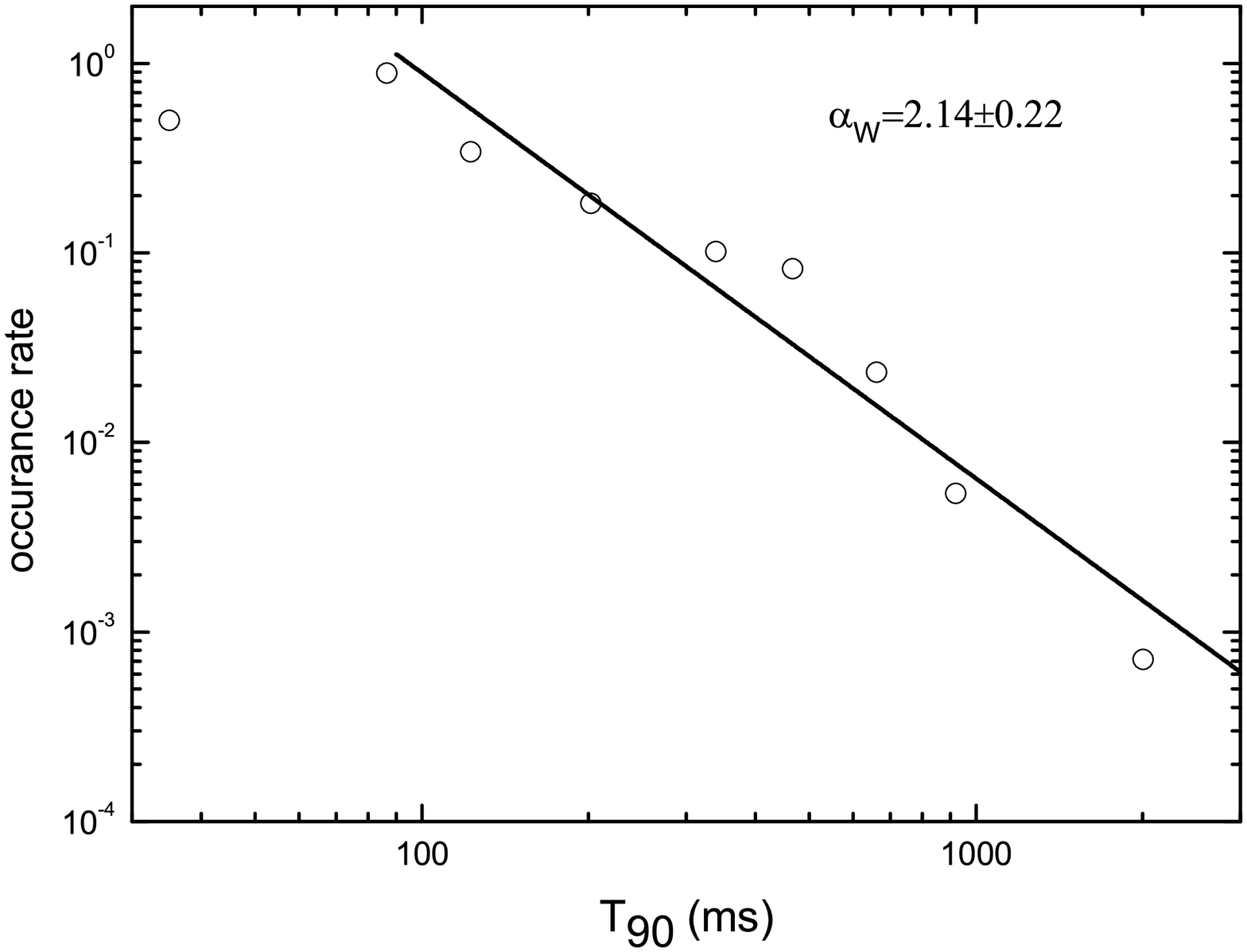}
\includegraphics[width=0.4\textwidth]{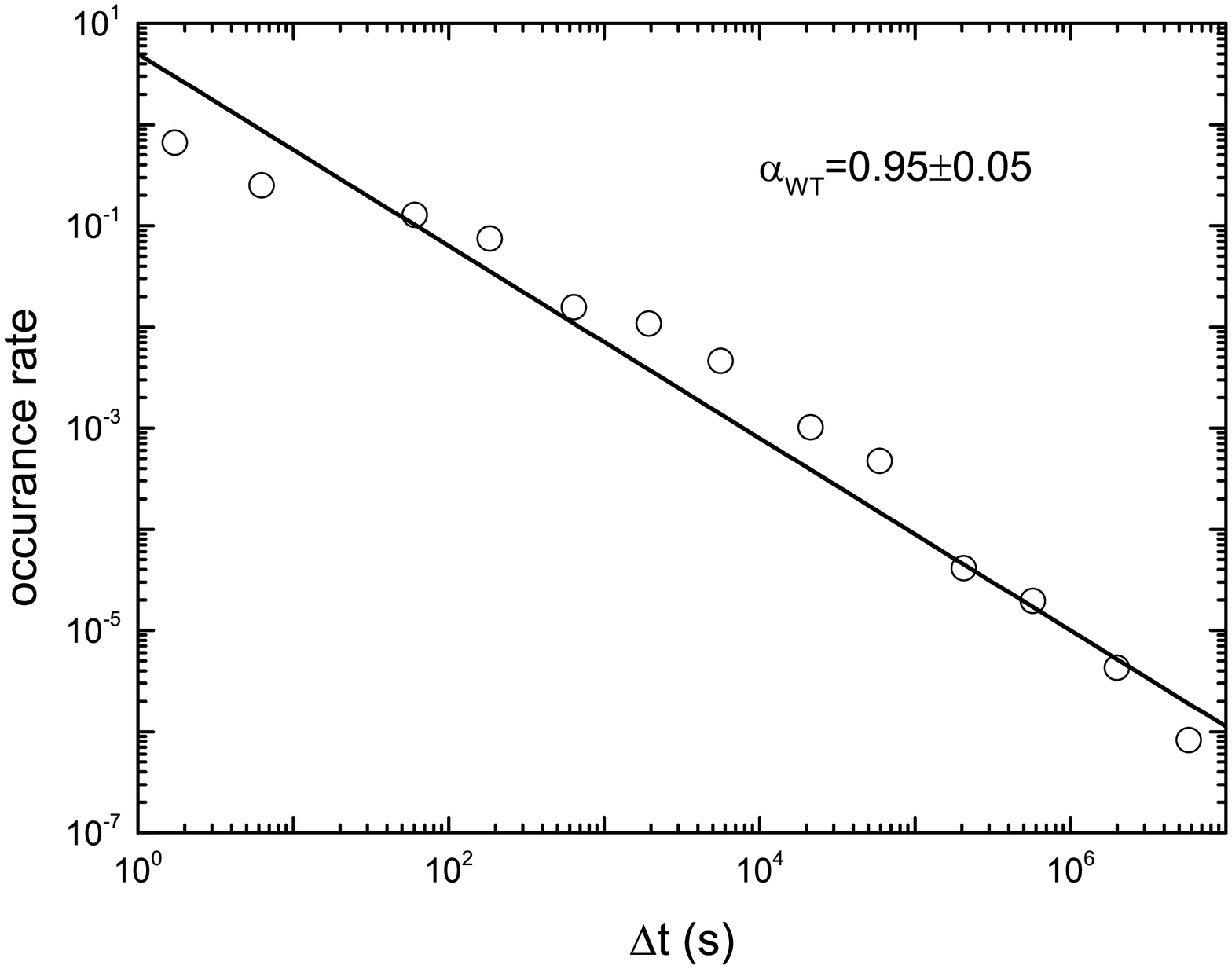}
\caption{The differential distributions of duration (left panel) and
waiting time (right panel) for SGR 1806-20, respectively. The
best-fitting power-law indices are $\alpha_{W}=2.14\pm0.22$ and
$\alpha_{WT}=0.95\pm0.05$, respectively. }
\end{figure*}

\end{document}